\documentclass[11pt,a4paper]{article}

\usepackage[T1]{fontenc}
\usepackage{lmodern}
\usepackage{amsmath,amssymb,geometry}
\usepackage{graphicx}

\usepackage{xcolor}

\usepackage[
    colorlinks=true,
    linkcolor=black,
    citecolor=blue,
    urlcolor=blue
]{hyperref}

\title{Microscopic Origin of the Equation of Motion for Open Thermodynamic Systems}

\author{E. Aydiner\\[6pt]
\small Department of Physics, İstanbul University, 
\small İstanbul 34134, Türkiye\\[4pt]
\small \texttt{ekrem.aydiner@istanbul.edu.tr}
}

\date{October 29, 2026, Ankara}

\begin{document}
\maketitle

\begin{abstract}
In my previous work (arXiv:2609.07810) \cite{Aydiner_2026}, I derived a second-order
equation of motion for open thermodynamic
systems from the First Law of Thermodynamics
and established its dynamical equivalence
with mechanical oscillators and electrical
RLC circuits.
In the present study, I independently
recover the same mathematical structure
of the equation of motion from
a mesoscopic model of particle transfer
between two reservoirs connected by a
channel. Starting from collective momentum
balance, particle conservation, and
reservoir chemical-potential response,
I derive the inertial--dissipative--restoring
dynamics and obtain explicit expressions
for its effective coefficients in terms
of the physical parameters of the system.
The resulting equation predicts distinct
damping regimes, characteristic relaxation
time scales, and resonance under periodic
driving. The natural frequency and resonance
conditions are explicitly determined by
the physical parameters of the particle-transfer
system. In the slow-transport limit,
the model recovers the conventional
Onsager-type linear flux--force relation.
The independent emergence of the same
dynamical structure from two distinct
physical approaches provides theoretical
support for the previously proposed
thermodynamic equation of motion.
By identifying the physical mechanisms
underlying its effective coefficients,
the present study strengthens the
physical foundations of the proposed
framework and supports the possibility
of describing open thermodynamic systems
within a common dynamical framework
shared with mechanical and electrical
systems.
\end{abstract}

\section{Introduction} \label{sec:intro}

In classical physics, the dynamical evolution of mechanical
and electrical systems can be described by explicit equations
of motion. A damped harmonic oscillator and a driven series
RLC circuit, for example, are governed by second-order
differential equations containing inertial, dissipative,
and restoring terms. These equations provide a unified
mathematical description of the dynamical response of
otherwise physically different systems.

Thermodynamics, however, is traditionally formulated
in terms of energy, entropy, and other state variables.
For an open thermodynamic system, the First Law of
Thermodynamics can be expressed through the fundamental
thermodynamic relation
\begin{equation}
dU = T\,dS - p\,dV + \mu\,dN,
\label{eq:firstlaw}
\end{equation}
where $U$, $S$, $V$, and $N$ denote internal energy,
entropy, volume, and particle number, respectively,
while $T$, $p$, and $\mu$ are the corresponding
thermodynamic variables. Although this relation
provides a compact description of thermodynamic
energy changes, it does not, by itself, constitute
a closed equation of motion governing the temporal
evolution of an open system.

The dynamical description of nonequilibrium
thermodynamic systems has been extensively investigated.
Onsager's linear irreversible thermodynamics
\cite{Onsager_1931_I,Onsager_1931_II},
extended irreversible thermodynamics \cite{Jou_2010},
and generalized hydrodynamic approaches \cite{Boon_1991}
provide important frameworks for describing transport,
relaxation, and dissipative processes. In particular,
finite relaxation times and inertial-like transport
responses are familiar features of generalized
transport theories \cite{Cattaneo_1948,Jou_2010}.
Nevertheless, obtaining a mechanical-like equation
of motion from the First Law of Thermodynamics,
under explicitly specified physical assumptions,
remains a distinct theoretical objective.

In my previous work \cite{Aydiner_2026},
I addressed this problem by starting from the
First Law of Thermodynamics for an open system.
Using a dimensional reduction approach, together
with constitutive assumptions motivated by extended
nonequilibrium thermodynamics, I transformed the
thermodynamic energy relation into a second-order
equation of motion. The resulting equation takes
the form
\begin{equation}
M\ddot{\phi}+R\dot{\phi}+K\phi=F(t),
\label{eq:prior}
\end{equation}
where $M$, $R$, and $K$ represent effective
thermodynamic inertia, dissipation, and restoring
response, respectively, and $F(t)$ denotes the
external driving term.

An important result of that work was the
identification of a common dynamical structure
underlying open thermodynamic systems, damped
mechanical oscillators, and driven series RLC
circuits. Despite their different physical
origins, these systems can be described by
mathematically equivalent equations of motion
within the assumptions of the formulation.
This correspondence provides a possible route
toward describing thermodynamic dynamics within
a mathematical framework familiar from
classical mechanics and electrical circuit theory.

However, the derivation presented in my previous
work raises two fundamental questions.
First, can the same equation of motion be obtained
independently from a concrete physical model
of transport, rather than directly from the
thermodynamic formulation?
Second, can the effective coefficients appearing in the equation be identified in terms of
particle properties, momentum relaxation,
reservoir thermodynamic response, and
the geometry of the system?

These questions constitute the central motivation of the present study.
To address them, I consider particle transfer
between two reservoirs connected by a narrow
channel. The source reservoir has an externally
controlled chemical potential, whereas the
receiving reservoir has a chemical potential
that changes with its particle number.
Starting from the collective momentum balance
of particles moving through the channel,
particle conservation, and the thermodynamic
response of the receiving reservoir, I derive
a second-order dynamical equation.

The resulting equation reproduces the
inertial--dissipative--restoring structure
previously obtained from the First Law of
Thermodynamics. More importantly, its coefficients
can be expressed in terms of identifiable
physical parameters: the particle mass,
channel length and cross-sectional area,
particle number density, momentum relaxation
rate, and chemical-potential susceptibility
of the receiving reservoir.

The purpose of the present work is therefore
twofold: to demonstrate that the previously
derived thermodynamic equation of motion
can emerge independently from a particle-transfer
model, and to clarify the physical origin
of its dynamical coefficients. The derivation
provides a mesoscopic realization based on
collective particle motion and reservoir
thermodynamics, rather than an exact reduction
from a microscopic many-particle Hamiltonian.

Beyond establishing the physical origin
of the effective dynamical coefficients,
the present formulation also provides
explicit predictions for the dynamical
behavior of particle transfer between
reservoirs. In particular, it identifies
the natural frequency, damping regimes,
and conditions for resonant response
in terms of the physical parameters
of the system. Furthermore, the resulting
particle-current dynamics connects
finite-time relaxation with the
Maxwell--Cattaneo description \cite{Cattaneo_1948} and
recovers the conventional Onsager
flux--force relation \cite{Onsager_1931_I} in the slow-transport
limit. These results provide additional
physical consequences of the proposed
thermodynamic equation of motion and
suggest possible routes toward its
experimental investigation.

The remainder of this paper is organized
as follows. First, I introduce the
particle-transfer model and establish
the collective momentum balance governing
transport through the connecting channel.
I then examine the chemical-potential
response of the receiving reservoir and
derive the resulting equation of motion,
identifying its effective dynamical
coefficients. Next, I investigate the
effective energy balance, dimensional
consistency, dynamical regimes,
characteristic time scales, and resonance
under periodic driving. Finally, I discuss
the physical implications of the results,
their relationship to the previously
proposed thermodynamic formulation,
and the main conclusions of the study.

\section{The Particle Transfer Model} \label{sec:Particle_transfer}

\begin{figure}[htbp]
\centering
\includegraphics[width=0.78\linewidth]{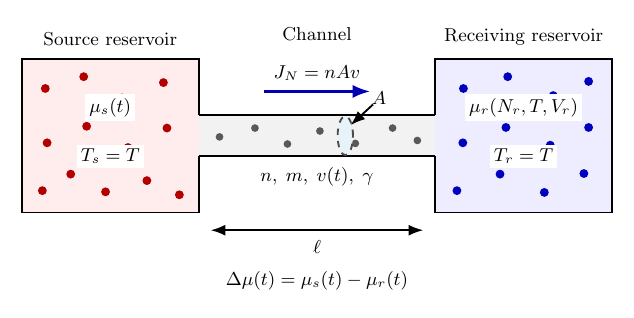}
\caption{Schematic representation of the isothermal
particle-transfer model. A source reservoir with
externally controlled chemical potential $\mu_s(t)$
is connected to a finite receiving reservoir with
chemical potential $\mu_r(N_r,T,V_r)$ through a
channel of length $\ell$ and cross-sectional area $A$.
The channel contains particles of mass $m$ and number
density $n$, moving with an average drift velocity
$v(t)$ and experiencing momentum relaxation
characterized by $\gamma$.
Although both reservoirs are maintained at the same
temperature, $T_s=T_r=T$, particle transfer is driven
by the chemical-potential difference
$\Delta\mu=\mu_s-\mu_r$.}
\label{fig:ata_fig1}
\end{figure}

The particle-transfer model considered in this work is
illustrated in Fig.~\ref{fig:ata_fig1}. The system consists
of a source reservoir and a finite receiving reservoir
connected by a narrow channel of length $\ell$ and
cross-sectional area $A$. Both reservoirs contain particles
of mass $m$, which can move through the connecting channel.
The source reservoir has an externally controlled chemical
potential $\mu_s(t)$, whereas the chemical potential of the
receiving reservoir, $\mu_r(N_r,T,V_r)$, depends on its
particle number $N_r$, temperature $T$, and volume $V_r$.

The two reservoirs are maintained at the same temperature,
$T_s=T_r=T$. Nevertheless, particle transfer can occur
because their chemical potentials are different. The
thermodynamic driving force is therefore the chemical-potential
difference
\begin{equation}
\Delta\mu(t)=\mu_s(t)-\mu_r(t).
\end{equation}
A positive chemical-potential difference favors particle
transfer from the source reservoir toward the receiving
reservoir. Thus, the model describes an isothermal
particle-transfer process driven by a chemical-potential
difference rather than a temperature gradient.

As indicated in Fig.~\ref{fig:ata_fig1}, the channel is
characterized by an approximately constant particle number
density $n$, a nearly uniform drift velocity $v(t)$,
and an effective momentum relaxation rate $\gamma>0$.
The parameter $\gamma$ represents the relaxation of
collective particle momentum due to dissipative processes
within the channel. The total number of particles contained
in the channel is approximately
\begin{equation}
N_c=nA\ell.
\end{equation}
The corresponding collective mass
of the moving particles is therefore
$mN_c=mnA\ell$.

The analysis assumes near-equilibrium conditions,
approximately uniform flow, and constant channel density.
The channel density is approximated
by the common equilibrium reference
density of the fluid, while deviations
from this reference state are assumed
to be sufficiently small.
Entrance and exit momentum-flux differences are neglected,
while interfacial pressure losses are assumed negligible
or incorporated into the effective dissipation.

To describe the particle-transfer dynamics, I introduce
the collective coordinate
\begin{equation}
\phi(t)=N_r(t)-N_0,
\end{equation}
where $N_0$ denotes the particle
number of the receiving reservoir
at the reference equilibrium state
about which the chemical potential
is subsequently linearized.

Applying particle conservation at the receiving
reservoir, the time derivative of the collective
coordinate gives
\begin{equation}
\dot\phi=\frac{dN_r}{dt}=nAv,
\end{equation}
where $n$, $A$, and $v$ denote the particle number
density, channel cross-sectional area, and
average drift velocity, respectively.
On the other hand, assuming a constant channel cross-sectional area
and approximately constant particle number density,
differentiation with respect to time yields
\begin{equation}
\ddot\phi=nA\dot v.
\label{eq:continuity}
\end{equation}
Here, $\dot\phi$ represents the net particle-transfer
rate into the receiving reservoir. Since $\phi$
is defined as a particle count, the corresponding
dynamical coefficients must be rescaled if the
transferred amount is expressed in moles.

\section{Channel Momentum Balance} \label{sec:Momentum_balance}

The channel contains $N_c=nA\ell$ particles
and has collective momentum
\begin{equation}
P=mnA\ell v.
\end{equation}
According to Newton's second law, the time
evolution of the collective momentum is
governed by the net force acting on the
particles in the channel,
\begin{equation}
\frac{dP}{dt}=F_{\mathrm{net}}.
\end{equation}
Under the stated control-volume approximations
\cite{Landau_1987},
the net force consists of the pressure-driven
force $A(p_s-p_r)$ and the dissipative
momentum-relaxation force $-m\gamma nA\ell v$.
Thus,
\begin{equation}
\frac{dP}{dt}
=
A(p_s-p_r)-m\gamma nA\ell v.
\end{equation}
Assuming approximately constant particle
number density and channel geometry,
the momentum balance becomes
\begin{equation}
mnA\ell\dot v
=
A(p_s-p_r)-m\gamma nA\ell v.
\label{eq:pressure}
\end{equation}
At constant temperature, the Gibbs--Duhem
relation for a single-component \cite{Callen_1985} fluid gives
\begin{equation}
dp=n\,d\mu.
\end{equation}
Consequently, the pressure difference
can be expressed as
\begin{equation}
p_s-p_r
=
\int_{\mu_r}^{\mu_s}n(\mu,T)\,d\mu.
\end{equation}
For small chemical-potential differences
around a common reference state,
the pressure difference becomes
\begin{equation}
p_s-p_r\simeq n(\mu_s-\mu_r),
\end{equation}
where the common reference density is
assumed to coincide with the approximately
constant particle density in the channel.

The chemical-potential difference
$\Delta\mu=\mu_s-\mu_r$ constitutes
the thermodynamic driving force
for particle transfer. In the channel
momentum balance, this imbalance
corresponds to the mechanical driving
force $F_{\mathrm{chem}}=nA\Delta\mu$.
When expressed in terms of the collective
particle-transfer coordinate, $\Delta\mu$
acts as the generalized driving force.

Substitution into Eq.~\eqref{eq:pressure}
gives
\begin{equation}
mnA\ell\dot v
=
nA(\mu_s-\mu_r)
-m\gamma nA\ell v.
\label{eq:momentum}
\end{equation}
The pressure difference associated with
the chemical-potential imbalance thus
provides the driving force for particle
transfer. This approximation applies
to the single-component, nearly isothermal
fluid considered here, with additional
interfacial forces and externally imposed
potentials neglected.

The dissipative contribution is described
by a phenomenological linear momentum-relaxation
term, where $\gamma$ denotes the effective
relaxation rate of the collective particle
momentum. This parameter represents
momentum-dissipating processes within
the channel and is not explicitly derived
from an underlying many-particle Hamiltonian.

Using Eq.~\eqref{eq:continuity},
the momentum balance can be expressed
in terms of the collective coordinate
$\phi$ as
\begin{equation}
\frac{m\ell}{nA}\ddot\phi
+
\frac{m\gamma\ell}{nA}\dot\phi
=
\mu_s-\mu_r.
\label{eq:preliminary}
\end{equation}
Equation~\eqref{eq:preliminary} establishes
a direct connection between the chemical-potential
difference and the collective dynamics
of particle transfer. The coefficients
of the second- and first-order time
derivatives are identified as
\begin{equation}
M=\frac{m\ell}{nA},
\qquad
R=\frac{m\gamma\ell}{nA}.
\label{eq:MR}
\end{equation}

Here, $M$ represents the effective inertia
associated with the collective momentum
of particles moving through the channel,
whereas $R$ characterizes dissipative
momentum relaxation.

An important consequence is that the
particle current does not respond
instantaneously to changes in the
chemical-potential difference.
Instead, its evolution is characterized
by a finite momentum relaxation time,
\begin{equation}
\tau=\frac{M}{R}=\frac{1}{\gamma}.
\end{equation}

Consequently, the particle-transfer process
already exhibits an inertial--dissipative
dynamical structure, independently of the
restoring response subsequently introduced
through the receiving reservoir.
This description extends the instantaneous
linear flux--force relation of conventional
irreversible thermodynamics
\cite{Onsager_1931_I}
by explicitly accounting for the finite
relaxation time of the particle current,
in a manner analogous to Maxwell--Cattaneo
transport models \cite{Cattaneo_1948}.

To complete the dynamical description,
I next examine the chemical-potential
response of the receiving reservoir
near equilibrium, which determines
the restoring coefficient $K$,
while the externally controlled
chemical potential of the source
reservoir provides the driving term $F(t)$.

\section{Chemical-Potential Driving Force and Restoring Response} \label{sec:Chemical_Pot}

Although the two reservoirs are maintained
at the same temperature, $T_s=T_r=T$,
particle transfer can occur because their
chemical potentials differ.

For an ideal gas, the chemical potential
is given by \cite{Callen_1985}
\begin{equation}
\mu(T,n_i)=\mu^\circ(T)+k_BT
\ln\left(\frac{n_i}{n^\circ}\right),
\qquad i=s,r,
\end{equation}
where $n_i$ denotes the particle number
density of the corresponding reservoir,
with $n_s$ and $n_r$ representing the
source and receiving reservoirs,
respectively, and $n^\circ$ is a reference
number density. The particle number density
in the connecting channel is denoted by
$n$ and is assumed to remain approximately
constant throughout the analysis.

At equal temperatures, the chemical-potential
difference becomes
\begin{equation}
\Delta\mu
=
\mu_s-\mu_r
=
k_BT\ln\left(\frac{n_s}{n_r}\right),
\end{equation}
where $n_s$ and $n_r$ denote the particle
number densities of the source and
receiving reservoirs, respectively.
Thus, when $n_s>n_r$, the chemical-potential
difference favors particle transfer from
the source toward the receiving reservoir,
even in the absence of a temperature gradient.

However, since the receiving reservoir is finite,
its chemical potential changes as particles
accumulate, giving rise to a restoring
response near the reference state.
Expanding the chemical potential of the
receiving reservoir around $N_r=N_0$
gives
\begin{equation}
\mu_r(N_r)
\simeq
\mu_0+K\phi,
\qquad
K=
\left(
\frac{\partial\mu_r}{\partial N_r}
\right)_{T,V_r}>0,
\label{eq:restoring}
\end{equation}
where $\phi=N_r-N_0$ and
$\mu_0=\mu_r(N_0)$.

For an ideal-gas receiving reservoir,
\begin{equation}
\mu_r
=
\mu^\circ(T)
+
k_BT\ln\left(
\frac{N_r}{V_r n^\circ}
\right),
\end{equation}
which yields
\begin{equation}
K=\frac{k_BT}{N_0}.
\end{equation}
This result differs from the effective
restoring coefficient $2k_BT/N_0$
obtained for two identical finite reservoirs,
each containing $N_0$ particles in the
reference state, exchanging particles
at fixed total particle number.
In the present model, the source
chemical potential is independently
controlled.

The external driving contribution
is defined by
\begin{equation}
F(t)=\mu_s(t)-\mu_0.
\label{eq:forcing}
\end{equation}
Consequently, the chemical-potential
difference becomes
\begin{equation}
\Delta\mu(t)=F(t)-K\phi(t).
\end{equation}
Substituting this expression into
Eq.~\eqref{eq:preliminary}, I obtain
the equation of motion
\begin{equation}
M\ddot\phi+R\dot\phi+K\phi=F(t).
\label{eq:eom}
\end{equation}
Remarkably, Eq.~\eqref{eq:eom} reproduces
the same second-order inertial--dissipative--restoring
structure that I previously derived
from the First Law of Thermodynamics
within the framework of extended
nonequilibrium thermodynamics
\cite{Aydiner_2026}.

The significance of this result lies
in the fundamentally different physical
starting points of the two derivations.
While the previous formulation was
obtained from a thermodynamic energy
balance, the present equation emerges
independently from Newton's second law,
particle conservation, and the
chemical-potential response of
the receiving reservoir.
Moreover, the present model explicitly
identifies the physical origins of
the three dynamical coefficients:
collective particle momentum gives
rise to effective inertia,
momentum relaxation produces
dissipation, and the chemical-potential
susceptibility of the receiving
reservoir generates the restoring
response. The external driving term
is determined by the controlled
chemical potential of the source
reservoir.

The independent recovery of the same
equation of motion provides concrete
mesoscopic support for the thermodynamic
dynamical framework proposed in my
previous work. It demonstrates that
the inertial--dissipative--restoring
structure is not restricted to its
original thermodynamic derivation,
but can also emerge directly from
the collective dynamics of
particle transfer.

\section{Effective Energy Balance} \label{sec:Energy_balance}

The equation of motion derived from the
mesoscopic particle-transfer model also
admits a physically meaningful effective
energy balance.

For constant coefficients, multiplying
Eq.~\eqref{eq:eom} by $\dot\phi$ yields
\begin{equation}
\frac{d}{dt}
\left(
\frac12 M\dot\phi^2
+
\frac12 K\phi^2
\right)
+
R\dot\phi^2
=
F(t)\dot\phi.
\label{eq:energy}
\end{equation}
The physical significance of this relation
can be established by identifying the
individual contributions directly from
the mesoscopic model.
The first quadratic term represents the
collective kinetic energy of particles
moving through the channel.
Indeed, using $\dot\phi=nAv$ and
$M=m\ell/(nA)$ gives
\begin{equation}
\frac12 M\dot\phi^2
=
\frac12 mnA\ell v^2.
\label{eq:kinetic_energy}
\end{equation}

Since $mnA\ell$ is the total mass
of particles contained in the channel,
Eq.~\eqref{eq:kinetic_energy} coincides
with their collective flow kinetic energy.
Thus, the effective inertial coefficient
$M$ is directly connected to the
mechanical energy of particle transport.

The restoring contribution has a
thermodynamic origin. At constant
temperature and volume, the chemical
potential of the receiving reservoir
satisfies
\begin{equation}
\mu_r=
\left(
\frac{\partial\mathcal F_r}
{\partial N_r}
\right)_{T,V_r},
\end{equation}
where $\mathcal F_r$ denotes its
Helmholtz free energy.
Expanding $\mathcal F_r$ around the
reference particle number $N_0$ gives
\begin{equation}
\mathcal F_r(N_0+\phi)
=
\mathcal F_r(N_0)
+
\mu_0\phi
+
\frac12 K\phi^2
+
O(\phi^3),
\end{equation}
where
\begin{equation}
K=
\left(
\frac{\partial\mu_r}
{\partial N_r}
\right)_{T,V_r,N_r=N_0}.
\end{equation}
Consequently, the excess free energy
relative to the reference chemical
potential is
\begin{equation}
\begin{aligned}
\Delta\mathcal F_{\mathrm{ex}}
&=
\mathcal F_r(N_0+\phi)
-
\mathcal F_r(N_0)
-
\mu_0\phi
\\
&\simeq
\frac12 K\phi^2.
\end{aligned}
\label{eq:free_energy}
\end{equation}

The effective stored energy can therefore
be written, to quadratic order in the
collective variables, as
\begin{equation}
E_{\mathrm{eff}}
=
\frac12 M\dot\phi^2
+
\frac12 K\phi^2.
\label{eq:effective_energy}
\end{equation}
This expression establishes a direct
connection between the two conservative
contributions of the equation of motion
and their physical energy counterparts:
collective kinetic energy in the channel
and excess Helmholtz free energy in
the receiving reservoir.

The dissipative contribution is
\begin{equation}
P_{\mathrm{diss}}
=
R\dot\phi^2
\geq 0.
\end{equation}
It represents the rate at which
collective flow energy is dissipated
through momentum relaxation.

Under isothermal conditions, assuming
that the dissipated energy is thermalized
at temperature $T$, the corresponding
entropy production rate is
\begin{equation}
\dot S_{\mathrm i}
=
\frac{R\dot\phi^2}{T}
\geq 0.
\label{eq:entropy_production}
\end{equation}
This result is consistent with the
Second Law of Thermodynamics \cite{deGroot_1984} and
identifies the nonnegative entropy
production associated with the
dissipative mechanism represented
in the present model.

The term $F(t)\dot\phi$ represents
the generalized power supplied by
the externally controlled source
chemical potential, measured relative
to the reference chemical potential
$\mu_0$.
Accordingly, Eq.~\eqref{eq:energy}
takes the form
\begin{equation}
\frac{dE_{\mathrm{eff}}}{dt}
+
P_{\mathrm{diss}}
=
P_{\mathrm{in}},
\label{eq:energy_balance}
\end{equation}
where
\begin{equation}
P_{\mathrm{in}}=F(t)\dot\phi.
\end{equation}
Equation~\eqref{eq:energy_balance}
expresses the balance between
effective energy storage,
irreversible dissipation, and
external chemical-potential driving.
Importantly, this energy balance
describes deviations around a
thermodynamic reference state.
It does not represent the complete
energy balance of the combined
reservoir--channel system, which
would also include background
chemical and thermal contributions.
The significance of the present
result extends beyond the formal
recovery of the same second-order
equation of motion obtained in
my previous thermodynamic formulation
\cite{Aydiner_2026}.
The mesoscopic model provides
explicit physical interpretations
of the inertial, dissipative,
and restoring contributions,
and demonstrates that the resulting
dynamics is compatible with an
effective energy balance.
Thus, the present derivation
establishes a concrete connection
between collective particle-transfer
dynamics and the energy-based
thermodynamic framework developed
in my previous work.
While this correspondence does
not by itself establish quantitative
identity between the coefficients
of the two formulations, it provides
additional physical support for
the proposed inertial--dissipative--
restoring dynamical structure.

\section{Dimensional Consistency}
\label{sec:Dimension}

An important question arising from the
present mesoscopic derivation is whether
the dynamical coefficients obtained
from collective particle momentum,
momentum relaxation, and reservoir
susceptibility form a dimensionally
consistent equation of motion.
This question is particularly relevant
because the present formulation
recovers the same second-order
inertial--dissipative--restoring
structure previously derived from
the First Law of Thermodynamics
\cite{Aydiner_2026}, although the
two approaches employ different
physical starting points and
definitions of the collective coordinate.
The dimensional analysis therefore
serves two purposes: to verify the
internal consistency of the mesoscopic
equation and its associated energy
balance, and to clarify the dimensional
correspondence between the present
model and the previous thermodynamic
formulation.

Since the collective coordinate
$\phi=N_r-N_0$ represents a particle
count, it is dimensionless. Therefore,
\begin{equation}
[\phi]=1,\qquad
[\dot{\phi}]=\mathrm{s^{-1}},\qquad
[\ddot{\phi}]=\mathrm{s^{-2}}.
\end{equation}
The particle mass $m$, channel length
$\ell$, cross-sectional area $A$,
number density $n$, and momentum
relaxation rate $\gamma$ have dimensions
\begin{equation}
[m]=\mathrm{kg},\quad
[\ell]=\mathrm{m},\quad
[A]=\mathrm{m^2},\quad
[n]=\mathrm{m^{-3}},\quad
[\gamma]=\mathrm{s^{-1}}.
\end{equation}
For the effective inertia,
\begin{equation}
[M]
=
\left[\frac{m\ell}{nA}\right]
=
\frac{\mathrm{kg\,m}}
{\mathrm{m^{-3}m^2}}
=
\mathrm{kg\,m^2}
=
\mathrm{J\,s^2}.
\end{equation}
Similarly, the effective resistance
has dimensions
\begin{equation}
[R]
=
\left[\frac{m\gamma\ell}{nA}\right]
=
\mathrm{kg\,m^2\,s^{-1}}
=
\mathrm{J\,s}.
\end{equation}
Since the chemical potential represents
energy per particle and the particle
number is dimensionless, the restoring
coefficient satisfies
\begin{equation}
[K]
=
\left[
\left(
\frac{\partial\mu_r}{\partial N_r}
\right)_{T,V_r}
\right]
=
\mathrm{J}.
\end{equation}
The external driving term is defined
through a chemical-potential difference
and therefore has dimensions
\begin{equation}
[F(t)]
=
[\mu_s-\mu_0]
=
\mathrm{J}.
\end{equation}

Consequently, all terms in the equation
of motion have identical dimensions:
\begin{equation}
[M\ddot{\phi}]
=
[R\dot{\phi}]
=
[K\phi]
=
[F(t)]
=
\mathrm{J}.
\end{equation}
Thus, the derived equation of motion
is dimensionally consistent.
The driving term $F(t)$ represents
a generalized force conjugate to
the dimensionless particle-transfer
coordinate $\phi$, rather than a
mechanical force measured in Newtons.

The dimensional consistency also
extends to the effective energy
balance derived in the preceding
section. In particular,
\begin{equation}
[E_{\mathrm{eff}}]=\mathrm{J},
\qquad
[P_{\mathrm{diss}}]
=
[P_{\mathrm{in}}]
=
\mathrm{J\,s^{-1}}.
\end{equation}
These results confirm that the
inertial, dissipative, restoring,
and driving contributions possess
mutually consistent dimensions,
both in the equation of motion
and in its associated energy balance.

It is important to emphasize that
the dimensional structure of the
equation of motion depends on the
physical definition of the collective
coordinate. In my previous work
\cite{Aydiner_2026}, the collective
coordinate was defined as the
transferred amount of substance
measured in moles. Accordingly,
the dynamical coefficients had
dimensions
\begin{equation}
[M]=\mathrm{J\,s^2\,mol^{-2}},
\qquad
[R]=\mathrm{J\,s\,mol^{-2}},
\qquad
[K]=\mathrm{J\,mol^{-2}}.
\end{equation}
Consequently, each term in the
previous equation of motion had
dimensions of $\mathrm{J\,mol^{-1}}$.

In contrast, the present formulation
uses the particle-number coordinate
$\phi=N_r-N_0$, which is dimensionless.
Accordingly, all terms in the equation
of motion have dimensions of energy,
$\mathrm{J}$. When the two coordinates describe
the same particle-transfer process,
they are related through the
Avogadro constant,
\begin{equation}
\phi_{\mathrm{particle}}
=
N_{\mathrm A}\phi_{\mathrm{mol}},
\end{equation}
where $N_{\mathrm A}$ has dimensions
of $\mathrm{mol^{-1}}$.
The corresponding generalized forces
and dynamical coefficients must
transform consistently with this
change of coordinate.
Thus, the difference in dimensional
structure between the two formulations
does not imply a physical or
mathematical inconsistency.
It reflects the different definitions
of the collective coordinate.
Nevertheless, dimensional consistency
alone does not establish quantitative
equality of the effective coefficients
obtained from the two derivations.

The dimensional analysis establishes
that the effective dynamical coefficients
derived from distinct physical mechanisms
are mutually consistent and generate
a well-defined equation of motion
and associated energy balance.
Furthermore, the apparent dimensional
differences between the present
mesoscopic formulation and the
previous thermodynamic derivation
are consistent with their different
definitions of the collective coordinate.
These findings provide an additional
internal consistency check for the
mesoscopic realization of the
previously proposed thermodynamic
equation of motion. Although
dimensional consistency alone
does not establish quantitative
equivalence of the two formulations,
it removes a potential dimensional
obstacle to their comparison and
supports the physical coherence
of the proposed dynamical framework.

\section{Dynamical Regimes, Characteristic Time Scales,
and Resonance}
\label{sec:dynamical_regimes}

The equation of motion derived in the preceding
sections provides a framework for investigating
the characteristic time scales, dynamical regimes,
and frequency-dependent response of particle
transfer between thermodynamic reservoirs.
The competition between collective inertia,
momentum relaxation, and the restoring response
of the receiving reservoir determines whether
the system exhibits oscillatory or nonoscillatory
relaxation. Under time-dependent chemical-potential
driving, the same dynamical structure also permits
resonant particle-transfer responses.
An important objective is therefore to determine
the conditions under which such behavior emerges
from the microscopic and geometric parameters
of the model.

\subsection{Free dynamics and characteristic frequencies}

For constant coefficients and in the absence
of external driving, the equation of motion
reduces to the standard damped-oscillator
form \cite{Landau_1976},
\begin{equation}
M\ddot{\phi}+R\dot{\phi}+K\phi=0.
\label{eq:free_dynamics}
\end{equation}
Seeking solutions of the form
$\phi(t)\propto e^{\lambda t}$ gives
the characteristic equation
\begin{equation}
M\lambda^2+R\lambda+K=0.
\label{eq:characteristic}
\end{equation}
The corresponding eigenvalues are
\begin{equation}
\lambda_{\pm}
=
-\frac{R}{2M}
\pm
\sqrt{
\left(\frac{R}{2M}\right)^2
-
\frac{K}{M}
}.
\label{eq:eigenvalues}
\end{equation}
The natural angular frequency of the
undamped system is
\begin{equation}
\omega_0
=
\sqrt{\frac{K}{M}}.
\label{eq:natural_frequency}
\end{equation}
Using the microscopic expressions
\begin{equation}
M=\frac{m\ell}{nA},
\qquad
K=
\left(
\frac{\partial\mu_r}{\partial N_r}
\right)_{T,V_r},
\end{equation}
one obtains
\begin{equation}
\omega_0
=
\sqrt{
\frac{nA}{m\ell}
\left(
\frac{\partial\mu_r}{\partial N_r}
\right)_{T,V_r}
}.
\label{eq:frequency_general}
\end{equation}

For an ideal-gas receiving reservoir,
$K=k_BT/N_0$, and therefore
\begin{equation}
\omega_0
=
\sqrt{
\frac{nAk_BT}{m\ell N_0}
}.
\label{eq:frequency_ideal}
\end{equation}
These expressions establish a direct
connection between the characteristic
frequency and the microscopic,
geometric, and thermodynamic properties
of the particle-transfer system.
In particular, the natural frequency
depends on the particle mass,
channel geometry, temperature,
and particle-number susceptibility
of the receiving reservoir.

The damping ratio is defined as
\begin{equation}
\zeta
=
\frac{R}{2\sqrt{MK}}.
\label{eq:damping_ratio}
\end{equation}
Since the present model gives
$R=\gamma M$, this becomes
\begin{equation}
\zeta
=
\frac{\gamma}{2\omega_0}.
\label{eq:damping_microscopic}
\end{equation}
The dynamical behavior is therefore
controlled by the competition between
the momentum relaxation rate $\gamma$
and the natural frequency $\omega_0$.

For $\gamma<2\omega_0$, the eigenvalues
are complex and the system exhibits
underdamped oscillations:
\begin{equation}
\phi(t)
=
e^{-\gamma t/2}
\left[
C_1\cos(\omega_d t)
+
C_2\sin(\omega_d t)
\right],
\label{eq:underdamped}
\end{equation}
where
\begin{equation}
\omega_d
=
\sqrt{
\omega_0^2-\frac{\gamma^2}{4}
}.
\label{eq:damped_frequency}
\end{equation}
The oscillation amplitude decays
with the characteristic time scale
\begin{equation}
\tau_{\mathrm{amp}}
=
\frac{2}{\gamma}.
\end{equation}
Similarly, for $\gamma=2\omega_0$, the system
is critically damped:
\begin{equation}
\phi(t)
=
(C_1+C_2t)e^{-\gamma t/2}.
\label{eq:critical_damping}
\end{equation}
and for $\gamma>2\omega_0$, both eigenvalues
are real and negative, and the system
exhibits overdamped relaxation:
\begin{equation}
\phi(t)
=
C_1e^{\lambda_+t}
+
C_2e^{\lambda_-t}.
\label{eq:overdamped}
\end{equation}
Thus, the existence of a second-order
equation of motion does not necessarily
imply oscillatory behavior.
Oscillations occur only when the
restoring response is sufficiently
strong relative to momentum relaxation.

\subsection{Periodic driving and resonant response}

A particularly important consequence
of the derived equation of motion is
its response to periodic variations
of the source chemical potential. The steady-state response can be analyzed
using the standard theory of periodically
driven damped oscillators
\cite{Landau_1976}.

Consider a harmonic driving term
\begin{equation}
F(t)=F_0\cos(\Omega t),
\label{eq:harmonic_driving}
\end{equation}
where $F_0$ is the driving amplitude
and $\Omega$ is the externally imposed
angular frequency. Thus, the equation of motion becomes
\begin{equation}
M\ddot{\phi}
+
R\dot{\phi}
+
K\phi
=
F_0\cos(\Omega t).
\label{eq:forced_dynamics}
\end{equation}
After the transient contribution
has decayed, the stationary response
can be written as
\begin{equation}
\phi(t)
=
\Phi(\Omega)
\cos[\Omega t-\delta(\Omega)].
\label{eq:stationary_response}
\end{equation}
On the other hand, the frequency-dependent amplitude is
\begin{equation}
\Phi(\Omega)
=
\frac{F_0}{
\sqrt{
(K-M\Omega^2)^2
+
(R\Omega)^2
}
}.
\label{eq:response_amplitude}
\end{equation}
Equivalently, introducing the
complex susceptibility
\begin{equation}
\chi(\Omega)
=
\frac{1}{
K-M\Omega^2+iR\Omega
},
\label{eq:susceptibility}
\end{equation}
the response amplitude is
\begin{equation}
\Phi(\Omega)
=
F_0|\chi(\Omega)|.
\end{equation}
Here, the phase lag is determined by
\begin{equation}
\delta(\Omega)
=
\operatorname{atan2}
\left(
R\Omega,\,
K-M\Omega^2
\right).
\label{eq:phase_lag}
\end{equation}
The frequency dependence of
$\Phi(\Omega)$ reveals the possibility
of a resonant enhancement of the
particle-number response.

Consequently, maximizing the amplitude with respect
to the driving frequency gives
\begin{equation}
\Omega_{\mathrm{res}}^2
=
\frac{K}{M}
-
\frac{R^2}{2M^2}.
\label{eq:resonance_general}
\end{equation}
and using $R=\gamma M$, the resonance
frequency becomes
\begin{equation}
\Omega_{\mathrm{res}}
=
\sqrt{
\omega_0^2-\frac{\gamma^2}{2}
}.
\label{eq:resonance_frequency}
\end{equation}
Here, a nonzero resonance frequency exists
only when
\begin{equation}
\gamma<\sqrt{2}\,\omega_0.
\label{eq:resonance_condition}
\end{equation}
This condition is more restrictive
than the condition for underdamped
free oscillations,
$\gamma<2\omega_0$.

Consequently, an underdamped system
does not necessarily exhibit a
finite-frequency maximum in its
particle-number response.
Hence, substituting the microscopic expression
for $\omega_0$ gives
\begin{equation}
\Omega_{\mathrm{res}}
=
\sqrt{
\frac{nA}{m\ell}
\left(
\frac{\partial\mu_r}{\partial N_r}
\right)_{T,V_r}
-
\frac{\gamma^2}{2}
}.
\label{eq:resonance_microscopic}
\end{equation}
For an ideal-gas receiving reservoir,
this reduces to
\begin{equation}
\Omega_{\mathrm{res}}
=
\sqrt{
\frac{nAk_BT}{m\ell N_0}
-
\frac{\gamma^2}{2}
}.
\label{eq:resonance_ideal}
\end{equation}
where, the corresponding resonance condition is
\begin{equation}
\gamma^2
<
\frac{2nAk_BT}{m\ell N_0}.
\label{eq:resonance_ideal_condition}
\end{equation}
These expressions identify the
microscopic and thermodynamic parameters
that control the existence and position
of the resonance peak.

In the weak-damping limit,
$\gamma\ll\omega_0$, the resonance
frequency approaches the natural
frequency,
\begin{equation}
\Omega_{\mathrm{res}}
\simeq\omega_0.
\end{equation}
At the resonance frequency, the
particle-number response reaches
the maximum amplitude
\begin{equation}
\Phi_{\mathrm{max}}
=
\frac{F_0}{
M\gamma
\sqrt{
\omega_0^2-\gamma^2/4
}
},
\label{eq:maximum_amplitude}
\end{equation}
provided that
$\gamma<\sqrt{2}\omega_0$.

The resonant response therefore
reflects the combined effects of
collective particle inertia,
reservoir susceptibility, and
momentum relaxation.
In particular, the restoring
coefficient $K$ is determined
by the thermodynamic response
of the receiving reservoir,
while the effective inertia $M$
is determined by the collective
momentum of particles in the channel.
The resonance is consequently
associated with the coupled
mechanical and thermodynamic
properties of the particle-transfer
system.

\subsection{Particle-current response}

The particle current is defined as
\begin{equation}
J_N(t)=\dot{\phi}(t).
\end{equation}
For harmonic driving, the stationary
current amplitude is
\begin{equation}
J_0(\Omega)
=
\Omega\Phi(\Omega).
\end{equation}
Thus,
\begin{equation}
J_0(\Omega)
=
\frac{F_0\Omega}{
\sqrt{
(K-M\Omega^2)^2
+
(R\Omega)^2
}
}.
\label{eq:current_amplitude}
\end{equation}
An important distinction arises
between the particle-number response
and the particle-current response.
For positive constant $M$, $R$,
and $K$, the current amplitude
reaches its maximum at
\begin{equation}
\Omega_{J,\mathrm{max}}
=
\omega_0.
\label{eq:current_resonance}
\end{equation}
The maximum current amplitude is
\begin{equation}
J_{0,\mathrm{max}}
=
\frac{F_0}{R}.
\label{eq:maximum_current}
\end{equation}
Unlike the particle-number response,
the current amplitude possesses
a finite-frequency maximum for
any positive damping coefficient
within the present constant-coefficient
model.
For strong damping, however,
this maximum becomes broad and
does not necessarily represent
a sharply resolved resonance.
The distinction between the
two response functions is
physically significant.
The particle-number response
characterizes the accumulation
and depletion of particles in
the receiving reservoir, whereas
the current response characterizes
the rate of particle transfer
through the connecting channel.
These observables therefore provide
complementary signatures of the
same underlying dynamics.

\subsection{Characteristic time scales and
physical accessibility of resonance}

The preceding analysis establishes
the mathematical conditions for
resonance. Its physical realization,
however, also requires consistency
with the approximations used in
the mesoscopic derivation.

The momentum relaxation time is
\begin{equation}
\tau_m
=
\frac{M}{R}
=
\frac{1}{\gamma}.
\label{eq:momentum_relaxation_time}
\end{equation}
The natural dynamical time scale is
\begin{equation}
\tau_0
=
\frac{1}{\omega_0}.
\label{eq:natural_timescale}
\end{equation}
The ratio
\begin{equation}
\frac{\omega_0}{\gamma}
=
\omega_0\tau_m
\label{eq:timescale_ratio}
\end{equation}
measures the relative importance
of collective inertia and
momentum relaxation.
The condition for a finite-frequency
maximum in the particle-number
response becomes
\begin{equation}
\omega_0\tau_m
>
\frac{1}{\sqrt{2}}.
\label{eq:timescale_resonance}
\end{equation}
Thus, the momentum relaxation
time must be sufficiently long
relative to the characteristic
dynamical time scale for a
particle-number resonance peak
to occur.
Here, a further consideration concerns
the assumption of approximately
uniform channel flow and density.
If pressure or density disturbances
propagate through the channel with
a characteristic speed $c_s$,
the corresponding propagation time is
\begin{equation}
\tau_{\mathrm{prop}}
=
\frac{\ell}{c_s}.
\label{eq:propagation_time}
\end{equation}
A sufficient time-scale separation
for treating the channel response
as approximately spatially uniform is
\begin{equation}
\Omega\tau_{\mathrm{prop}}
\ll1.
\label{eq:uniform_flow_condition}
\end{equation}
Near the natural frequency, this
condition becomes
\begin{equation}
\omega_0\frac{\ell}{c_s}
\ll1.
\label{eq:frequency_validity}
\end{equation}

Combining this requirement with
the particle-number resonance
condition identifies a possible
regime of consistent resonant dynamics:
\begin{equation}
\gamma
<
\sqrt{2}\,\omega_0,
\qquad
\omega_0
\ll
\frac{c_s}{\ell}.
\label{eq:resonance_validity}
\end{equation}
These inequalities are sufficient
time-scale criteria within a
quasi-uniform channel description,
rather than universal requirements
for resonance in all particle-transfer
systems.
The driving amplitude must also
remain sufficiently small for
the linear chemical-potential
response and approximately constant
transport coefficients to remain valid.
For the receiving reservoir, this
requires the particle-number
deviation to remain small compared
with the scale over which its
chemical-potential susceptibility
changes appreciably.
In particular, for an ideal-gas
reservoir, a natural requirement is
\begin{equation}
|\phi(t)|\ll N_0.
\label{eq:linear_response_condition}
\end{equation}

Under periodic driving, the
corresponding condition is
\begin{equation}
\Phi(\Omega)\ll N_0.
\label{eq:amplitude_validity}
\end{equation}
These restrictions identify the
parameter regime in which the
resonance predicted by the reduced
equation can be interpreted
consistently within the
underlying mesoscopic model.

\subsection{Connection with Maxwell--Cattaneo
and Onsager Dynamics}

The present formulation establishes
a direct connection with two
fundamental descriptions of
irreversible transport:
Maxwell--Cattaneo relaxation \cite{Cattaneo_1948}
and Onsager's linear
flux--force relations \cite{Onsager_1931_I}.
The present formulation establishes a direct connection with two fundamental descriptions of irreversible transport: Maxwell--Cattaneo flux relaxation and Onsager's linear flux--force relations. The former describes finite-time current relaxation, whereas the latter emerges as the slow-transport limit of the present mesoscopic dynamics.

Before introducing the restoring
response of the receiving reservoir,
the particle-transfer equation reads
\begin{equation}
M\ddot{\phi}+R\dot{\phi}
=
\Delta\mu.
\label{eq:dynamic_flux}
\end{equation}
Introducing the particle current
$J_N=\dot{\phi}$ gives
\begin{equation}
M\dot J_N+RJ_N=\Delta\mu.
\label{eq:current_relaxation}
\end{equation}
Equivalently,
\begin{equation}
\tau_m\dot J_N+J_N
=
\frac{\Delta\mu}{R},
\qquad
\tau_m=\frac{M}{R}=\frac{1}{\gamma}.
\label{eq:relaxation_equation}
\end{equation}
This equation has the mathematical
structure of a Maxwell--Cattaneo
flux-relaxation law
\cite{Cattaneo_1948,Joseph_1989},
in which the current responds to its
driving force over a finite
relaxation time.

In the present model, however,
the relaxation dynamics follows
directly from the collective
momentum balance of particles
moving through the channel,
rather than being introduced
as an independent phenomenological
constitutive assumption.
Furthermore, the effective
transport coefficients are
explicitly determined by the
particle properties and
channel geometry.

An important consequence is
that the conventional Onsager
description \cite{Onsager_1931_I} emerges naturally
in the slow-transport limit.
When the characteristic time
scale of the current variation
is much longer than
$\tau_m$, the inertial term
becomes negligible, yielding
\begin{equation}
J_N\simeq\frac{\Delta\mu}{R}.
\label{eq:onsager_limit}
\end{equation}

For an isothermal system,
the thermodynamic force
can be defined as
$X_N=\Delta\mu/T$.
The corresponding linear
flux--force relation becomes
\begin{equation}
J_N=L_{NN}X_N,
\qquad
L_{NN}=\frac{T}{R}.
\label{eq:onsager_coefficient}
\end{equation}
Thus, Onsager's linear
transport relation is recovered
as the low-frequency,
long-time limit of the
present inertial--dissipative
transport dynamics.

Since $R=\gamma M$,
the coefficients $M$ and $R$
are not independent.
The Onsager limit therefore
corresponds to negligible
current acceleration relative
to momentum relaxation,
rather than to an independent
reduction of $M$ at fixed $R$.

When the current-relaxation
equation is coupled to the
chemical-potential response
of the finite receiving reservoir,
\begin{equation}
\Delta\mu=F(t)-K\phi,
\end{equation}
the full equation of motion
is recovered:
\begin{equation}
M\ddot{\phi}
+
R\dot{\phi}
+
K\phi
=
F(t).
\end{equation}

This result reveals the
physical origin of the
oscillatory and resonant
behavior of the system.
Finite current relaxation,
arising from collective
particle inertia, combines
with the restoring response
of the reservoir to produce
a second-order dynamical
equation supporting damped
oscillations and resonance
under periodic driving.

The present model therefore
connects conventional
irreversible transport
to inertial thermodynamic
dynamics within a single
mesoscopic framework.
It also provides an independent
physical realization of
the second-order equation
of motion proposed in my
previous work
\cite{Aydiner_2026}.

\section{Discussion}
\label{sec:Disc}

In my previous work \cite{Aydiner_2026},
I proposed a second-order equation of motion
for open thermodynamic systems based on
the First Law of Thermodynamics and
established its dynamical equivalence
with mechanical oscillators and electrical
RLC circuits.
In the present study, I have independently
recovered the same equation of motion
from the collective dynamics of particle
transport between two reservoirs.
The convergence of these two distinct
theoretical approaches toward the same
inertial--dissipative--restoring structure
provides independent theoretical support
for my previously proposed thermodynamic
equation of motion.

The broader significance of these studies
may also be viewed in the context of
Prigogine's pioneering investigations
of irreversible processes and
nonequilibrium thermodynamics
\cite{Prigogine_1967,Nicolis_Prigogine_1977}.
While Prigogine's framework emphasized
entropy production, irreversible evolution,
and the emergence of dissipative structures,
my approach focuses on deriving an explicit
second-order equation of motion for
open thermodynamic systems and identifying
its connections with mechanical and
electrical dynamics.
The present mesoscopic derivation
provides a concrete physical realization
of this approach, suggesting a possible
route toward a common dynamical description
of thermodynamic and mechanical processes.

The significance of this result extends
beyond the mathematical equivalence
between thermodynamic, mechanical,
and electrical systems.
In the present work, I have identified
explicit physical origins for the
effective dynamical coefficients
in terms of collective particle inertia,
momentum relaxation, and reservoir
thermodynamic susceptibility.
The resulting mesoscopic realization
strengthens the physical foundation
of my earlier formulation and provides
further motivation for investigating
the connection between open thermodynamic
systems and the analytical framework
of classical mechanics, including
Lagrangian and Hamiltonian descriptions
under appropriate conditions.

An important theoretical question,
however, remains unresolved.
Although my two approaches lead to
the same mathematical equation,
the quantitative relationship between
their effective dynamical coefficients
has not yet been established.
In particular, the connection between
the thermodynamic inertia introduced
through nonequilibrium entropy corrections
in my previous work and the collective
mechanical inertia derived from particle
momentum in the present study requires
further investigation.
Establishing this relationship would
provide a deeper understanding of
the physical foundations of the
proposed thermodynamic dynamics.

An equally important direction is
the experimental investigation of
the predicted dynamical regimes,
particularly the frequency-dependent
response and resonance of particle
transfer between reservoirs.
The explicit resonance conditions
obtained in terms of particle mass,
channel geometry, temperature,
reservoir susceptibility, and
momentum relaxation provide
quantitative criteria for identifying
experimentally accessible parameter
regimes.
Such investigations would help
establish the range of validity,
physical generality, and experimental
relevance of the proposed equation
of motion.

\section{Conclusion}
\label{sec:Con}

In this study, I have independently
derived the second-order equation of
motion previously obtained in my
thermodynamic formulation
\cite{Aydiner_2026}, using a mesoscopic
model of particle transfer between
two reservoirs.
Starting from collective momentum
balance, particle conservation,
and reservoir chemical-potential
response, I have identified the
physical origins of effective
inertia, dissipation, and restoring
response.
This independent derivation provides
a concrete mesoscopic realization
of my previously proposed equation
of motion and strengthens its
theoretical foundations.

Beyond recovering the equation of
motion, I have established an
effective energy balance and
determined the conditions for
damped oscillations and resonant
particle-transfer dynamics.
The natural frequency, damping
regimes, and resonance conditions
are explicitly expressed in terms
of the physical parameters of
the system.
Furthermore, I have shown that
the particle-current dynamics
exhibits a Maxwell--Cattaneo-type
relaxation structure, while the
conventional Onsager flux--force
relation emerges in the
slow-transport limit.
These results connect collective
particle transport, irreversible
thermodynamics, and oscillatory
dynamics within a single
mesoscopic framework.

Taken together, my previous
thermodynamic derivation and the
present mesoscopic investigation
establish two distinct theoretical
routes to the same second-order
dynamical structure.
Although the quantitative
correspondence between their
effective coefficients remains
an open question, the present
results provide independent
theoretical support for the
proposed equation of motion
and demonstrate its realization
in a physically specified
particle-transfer system.
I regard these findings as a
further step toward developing
a common dynamical framework
connecting open thermodynamic
systems with mechanical and
electrical dynamics, and toward
exploring their formulation
within the broader analytical
structure of classical physics.

\bibliographystyle{unsrt}
\bibliography{ata}.  

\end{document}